\documentclass[prb,twocolumn,showpacs,aps,superscriptaddress,floatfix]{revtex4}
\usepackage{amsmath}
\usepackage{amssymb}
\usepackage{bm}
\usepackage{graphicx}
\usepackage{color}
\usepackage{hyperref}
\usepackage{verbatim}
\usepackage{float}

\begin{document}

\title{Thermodynamics of symmetric spin--orbital model: One- and two-dimensional cases}

\author{V.E. Valiulin}
\affiliation{Moscow Institute for Physics and Technology (State
University), Dolgoprudnyi, 141700 Russia}
\affiliation{National Research Center Kurchatov Institute, Moscow, 123182 Russia}

\author{A.V. Mikheyenkov}
\affiliation{Moscow Institute for Physics and Technology (State
University), Dolgoprudnyi, 141700 Russia}
\affiliation{National Research Center Kurchatov Institute, Moscow, 123182 Russia}
\affiliation{Institute for High Pressure Physics, Russian Academy of Sciences, Troitsk, Moscow, 108840 Russia}

\author{K.I. Kugel}
\affiliation{Institute for Theoretical and Applied Electrodynamics, Russian Academy of Sciences, Moscow, 125412 Russia}
\affiliation{National Research University Higher School of Economics, Moscow, 101000 Russia}

\author{A.F. Barabanov}
\affiliation{Institute for High Pressure Physics, Russian Academy of Sciences, Troitsk, Moscow, 108840 Russia}

\begin{abstract}
The specific heat and susceptibilities for the two- and one-dimensional spin--orbital models are calculated in the framework of a spherically symmetric self-consistent approach at different temperatures and relations between the parameters of the system. It is shown that even in the absence of the long-range spin and orbital order, the system exhibits the features in the behavior of thermodynamic characteristics, which are typical of those manifesting themselves at phase transitions. Such features are attributed to the quantum entanglement of the coupled spin and orbital degrees of freedom.
\end{abstract}


\date{\today}

\maketitle

\section{Introduction}
\label{Intro}

The problem concerning the formation of entangled quantum states plays an important role in many fields of physics being especially attractive in the research related to quantum computations and their possible implementations~\cite{Bengts06_Book,Benent07_Book}. The entanglement of states corresponding to different spin projections is most often considered. No less fascinating problems also arise in two-spin systems, where the entanglement is perceived as some correlated state of two spin variables~\cite{You15_NJP}.

The two-spin models themselves usually appear in the description of specific features of transition metal compounds with the coupled spin and orbital degrees of freedom; that is why such models are often referred to as spin--orbital ones~\cite{Kugel82_UFN,Oles09_APPA,Oles12_JPCM}. Unusual effects related to the spin--orbital correlations and the corresponding quantum entanglement are widely discussed in the current literature. In particular, the possibility of extraordinary spin--orbital quantum states and transitions between them was pointed out~\cite{Belemu17_PRB,Belemu18_NJP,Brzezi18_JSNM}.

It is interesting that an unusual behavior of spinorbital correlations can manifest itself also in systems without any spin or orbital order owing to their low dimensionality. For example, the symmetric two-dimensional model is characterized by vanishing spin--orbital correlations at certain threshold values of the temperature or parameters of the model~\cite{Kaga14_JL_R}. At the same time, the behavior of spin and orbital correlation functions themselves does not demonstrate any peculiarities. 

his work is aimed at finding out the thermodynamic manifestations of such unusual specific features of spin--orbital correlations in the systems under study. Below, we demonstrate that, although long-range order is absent because of the low dimensionality of the system, a steep threshold type increase in spinorbital correlations at a certain intersubsystem exchange coupling constant is accompanied by clearly pronounced features in the thermodynamic characteristics resembling those typical of a phase transition. We relate these features to the formation of an entangled spinorbital state.

We study the one-dimensional (1D) and two-dimensional (2D) cases using the spherically symmetric self-consistent approach, which provides quite reliable results for low-dimensional systems~ \cite{Kaga14_JL_R,Bara11_TMP_R}. Note that the specific features in the correlation functions and in the thermodynamic parameters manifest themselves in different ways depending on the dimensionality of the system under study. In particular, in the accepted approach, a reentrant (with respect to the temperature) transition to the state with nonzero intersubsystem correlations is possible in the 1D case in contrast to the 2D one.

\section{Model and method}
\label{Model}

We consider a symmetrical version of the quantum spin--orbital model (symmetrical Kugel--Khomskii model) on a square lattice (or a linear chain). The corresponding Hamiltonian has the form
\begin{equation*}
\widehat{\mathbf{H}} = J\sum_{<\mathbf{i},\mathbf{j}>}
{\mathbf{S}}_{\mathbf{i}}{\mathbf{S}}_{\mathbf{j}}+
I\sum_{<\mathbf{i},\mathbf{j}>}{\mathbf{T}}_{\mathbf{i}}
{\mathbf{T}}_{\mathbf{j}}
+K\sum_{<\mathbf{i},\mathbf{j}>}
\left({\mathbf{S}}_{\mathbf{i}}{\mathbf{S}}_{\mathbf{j}}\right)
\left( {\mathbf{T}}_{\mathbf{i}}{\mathbf{T }}_{\mathbf{j}}\right),
\end{equation*}
\begin{equation}
\widehat{\mathbf{H}}=\widehat{J}+\widehat{I}+\widehat{K} ,  \label{Hamilt}
\end{equation}
where $<\mathbf{i},\mathbf{j}>$ denotes the summation over the nearest neighbor bonds on the square lattice or linear chain; $\widehat{\mathbf{S}}_{\mathbf{i}}$ and $\widehat{\mathbf{T}}_{\mathbf{i}}$ are respectively the spin and pseudospin operators for $S=1/2$ and $T=1/2$, respectively (the latter operators correspond to the orbital degrees of freedom).

We study the case of antiferromagnetic (AFM) spin--spin and pseudospin--pseudospin interactions, $J=I>0$, as well as of the negative exchange coupling between subsystems, $K<0$  (below, we always use the energy units corresponding to $J=I=1$). In Ref.~\onlinecite{Pati98_PRL}, it was demonstrated within the symmetric spin--orbital model that the interplay of spin and pseudospin (orbital) degrees of freedom is the most clearly pronounced at such relation between the parameters and results in the  ``entanglement" of these degrees of freedom~\cite{You12_PRB,Lundgr12_PRB}.

According to the Mermin--Wagner theorem~\cite{Mermin66_PRL}, the long-range order is not possible for the decoupled subsystems ($K = 0$) in the 2D and 1D cases at any nonzero temperature $T \neq 0$. It is natural to suppose that switching on the intersubsystem coupling, $K \neq 0$, leads to an additional contribution of thermal and quantum fluctuations, and the long-range order is still missing. Below, we limit ourselves only to the case of nonzero temperatures.

Thus, we adopt the following assumptions:

\begin{itemize}
  \item (i) All lattice sites in the system under study are equivalent (the translational symmetry is not broken).
  \item (ii) All single-site averages are equal to zero (the long-range order is absent)
      \begin{equation}
\langle\widehat{{S}}_{\mathbf{i}}\rangle =\langle \widehat{{T}}_
{\mathbf{i}}\rangle =0 \label{site_aver}
\end{equation}
  \item (iii) Correlation functions involving different spin and pseudospin components ($\alpha \neq \beta$) also vanish (SU(2) symmetries in the spin and pseudospin spaces are not broken)
      \begin{equation}
\langle \widehat{S}_{\mathbf{i}}^{\alpha}\widehat{S}_{\mathbf{j}}^{\beta}\rangle=0, \quad \langle \widehat{T}_{\mathbf{i}}^{\alpha}\widehat{T}_{\mathbf{j}}^{\beta}\rangle=0, \quad \langle \widehat{S}_{\mathbf{i}}^{\alpha}\widehat{T}_{\mathbf{j}}^{\beta}\rangle=0 . \label{ini_corrs}
\end{equation}

\end{itemize}

Conditions (i)--(iii) are automatically satisfied in the quantum spherically symmetric self-consistent approach (see, e.g., Refs.~\onlinecite{Bara11_TMP_R,Kondo72_PTP,Shimah91_JPSJ}), which we use below.

Similarly to Ref.~\onlinecite{Kaga14_JL_R}, we consider the spin--spin and spin--pseudospin retarded Green's functions
\begin{equation}
G_{\mathbf{q}} = \left\langle S_{\mathbf{q}}^{z}\mid
S_{\mathbf{-q}}^{z}\right\rangle _{\omega } , \label{Gq0}
\end{equation}
\begin{equation}
R_{\mathbf{q}} = \left\langle T_{\mathbf{q}}^{z}\mid
S_{\mathbf{-q}}^{z}\right\rangle _{\omega } . \label{Rq0}
\end{equation}
Obviously, $\left\langle T_{\mathbf{q}}^{z}\mid T_{\mathbf{-q}}^{z}\right\rangle _{\omega} = \left\langle S_{\mathbf{q}}^{z}\mid S_{\mathbf{-q}}^{z}\right\rangle_{\omega}$   since we consider the symmetric case $I=J$.

The calculations employing the conventional spherically symmetric self-consistent scheme (see, e.g., Refs.~\onlinecite{Bara11_TMP_R,Kaga14_JL_R}) including the approximations for $\widehat{K}$ used in Ref.~\onlinecite{Kaga14_JL_R} and the dominant contribution of the on-site intersubsystem exchange coupling, result in the following expressions for $G_{\mathbf{q}}$ and $R_{\mathbf{q}}$:
\begin{equation}
G_{\mathbf{q}} = \frac{F_{ac}(\mathbf{q})}{\omega ^{2}-\omega
_{ac}^{2}(\mathbf{q})}+\frac{F_{opt}(\mathbf{q})}{\omega ^{2}-\omega
_{opt}^{2}(\mathbf{q})} , \label{Gq}
\end{equation}
\begin{equation}
R_{\mathbf{q}} = \frac{F_{ac}(\mathbf{q})}{\omega ^{2}-\omega
_{ac}^{2}(\mathbf{q})}-\frac{F_{opt}(\mathbf{q})}{\omega ^{2}-\omega
_{opt}^{2}(\mathbf{q})} , \label{Rq}
\end{equation}

Rather lengthy expressions for the numerators in the Green's functions and for the spectra corresponding to the optical and acoustic branches of excitations are given in Appendix.

Expressions \eqref{Gq} and \eqref{Rq} involve spin--spin correlation functions. In the two-dimensional case,
\begin{equation}
c_{r}=\langle \widehat{S}_{\mathbf{i}}^{z}\widehat{S}_{\mathbf{i+r}}^{z}\rangle,
\quad r = g, d, 2g
\end{equation}
are the spin--spin correlation functions for the first (side of the square),  $c_{g} \equiv c_{1}$, second (diagonal), $c_{d} \equiv c_{2}$, and third (doubled side), $c_{2g} \equiv c_{3}$, nearest neighbors, respectively. In the one-dimensional case, the diagonal correlation functions are, of course, absent.

The on-site and intersite spin--pseudospin correlation functions are
\begin{equation}
m_{0}=\langle S_{\mathbf{i}}^{z}T_{\mathbf{i}}^{z}\rangle,\ \
m_{g}=\langle S_{\mathbf{i}}^{z}T_{\mathbf{i+g}}^{z}\rangle ,
\end{equation}
respectively.

In the further numerical procedure, all correlation functions $c_{r}$, $m_{0}$, and $m_{g}$ are self-consistently determined in terms of the Green's functions $G$ and $R$.

\section{2D case. Results and discussion}
\label{2D_results}

An important qualitative result for the 2D case has been already reported in Ref.~\cite{Kaga14_JL_R}. Namely, at a fixed temperature and a small intersubsystem coupling parameter $|K|$, the spin--pseudospin correlation functions $m_{0}$ and $m_{1}$ vanish. However, when $|K|$  increases to a threshold value $K_c$, the correlation functions $m_{0}$ and $m_{1}$ become nonzero and increase according to a power law. A similar situation occurs at fixed $K$ with the decrease in the temperature. The correlation functions $m_{0}$ and $m_{1}$ start to increase in a power-law manner below the threshold temperature $T_c$.

In both cases, the intrasubsystem (spin--spin) correlation functions $c_r$  change only slightly and do not exhibit any anomalies at the critical point. For the sake of convenience, the corresponding plots taken from Ref.~\onlinecite{Kaga14_JL_R} are shown in Appendix.

Naturally, the arising intersubsystem correlations should appreciably affect the thermodynamic characteristics. In the framework of our approach, the energy is completely determined by one- and two-site correlation functions. Therefore, one could expect that the steep increase in the spin--pseudospin correlation functions in the entangled state leads to anomalies of the specific heat.

\begin{figure}[H]
\begin{center}
\includegraphics[width=0.8\columnwidth]{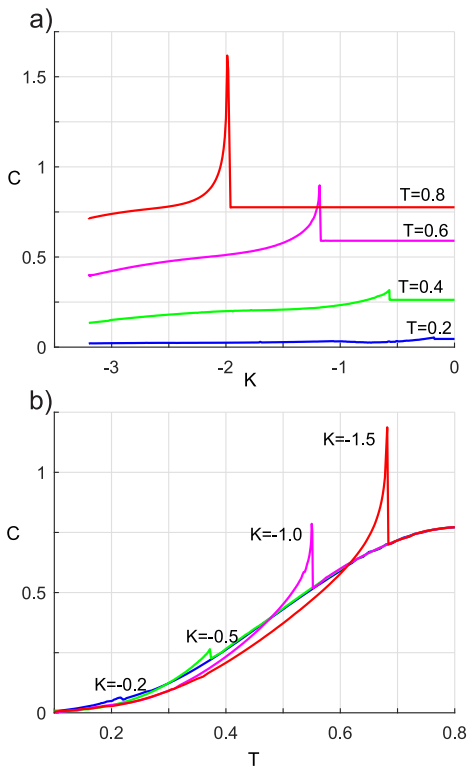}
\caption{\label{Fig1} Two-dimensional lattice. (a) Specific heat versus the intersubsystem exchange coupling constant $K$ at indicated temperatures $T$  given in different colors. (b) Temperature dependence of the specific heat at indicated exchange coupling constants $K$ given in different colors.}
\end{center}
\end{figure}

Figure~\ref{Fig1} illustrates the evolution of the specific heat (a) as a function of $K$ at fixed   and (b) as a function of $T$ at fixed $K$. As expected, at entering the range with intersubsystem correlations, the specific heat undergoes a jump. The higher the critical temperature $T_c$, the larger the jump. In Fig.~\ref{Fig1}b, we can also see that all curves have a common asymptotic behavior at high temperatures and $C \to 0$  at $T \to 0$  (the Nernst theorem is satisfied).

\begin{figure}[t]
\begin{center}
\includegraphics[width=0.8\columnwidth]{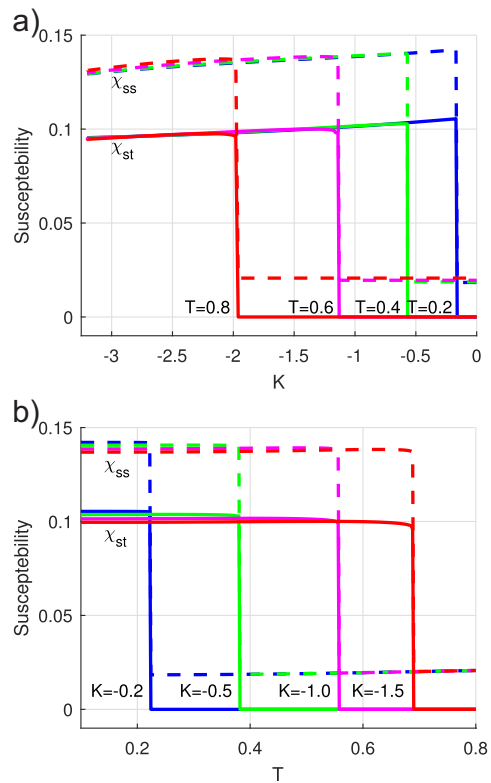}
\caption{\label{Fig2}
 Two-dimensional lattice. (a) Susceptibility versus the intersubsystem exchange coupling constant $K$ at indicated temperatures $T$   given in different colors. (b) Susceptibility versus temperature $T$ at indicated intersubsystem exchange coupling constants $K$ given in different colors. In both panels, dashed and solid lines denote the spin--spin susceptibility $\chi_{ss}$ and spin--pseudospin susceptibility $\chi_{st}$ , respectively.}
\end{center}
\end{figure}

Figure~\ref{Fig2} shows the behavior of spin--spin $\chi_{ss}$  and spin--pseudospin $\chi_{st}$  susceptibilities along the same paths in the phase diagram: (a) as functions of $K$ at fixed $T$  and (b) as functions of $T$ at fixed $K$. Similarly to the specific heat, the susceptibility exhibits a jump at critical points.

Outside the transition range, the susceptibility only slightly depends on $T$  and $K$. The behavior of the spin--spin susceptibility at small $|K|$ is in good agreement with that characteristic of the well-studied Heisenberg model~\cite{Shimah91_JPSJ}.

Thus, in the 2D case, i.e., for the square lattice, the state with entangled spin and pseudospin degrees of freedom arises at any temperature with the increase in the absolute value of the intersubsystem coupling constant $K$.

\section{1D case. Results and discussion}
\label{1D_results}

The calculations in the one-dimensional case are similar to those discussed above for the 2D case (see the corresponding expressions in Appendix).

Here, we adopt a mean-field approach implying the absence of decay of the spin excitations in the Green's functions given by Eqs.~\eqref{Gq} and \eqref{Rq}. Nevertheless, due to underlying conditions (i)--(iii) (see Section~\ref{Model}), such approach in the 1D case gives the results in good agreement with the exact solution or, in the situation with frustrations when the exact solution does not exist, with the numerical results for finite chains~\cite{Suzuki94_JPSJ,Junger04_PRB,Hartel08_PRB}.

\begin{figure}[t]
\begin{center}
\includegraphics[width=0.8\columnwidth]{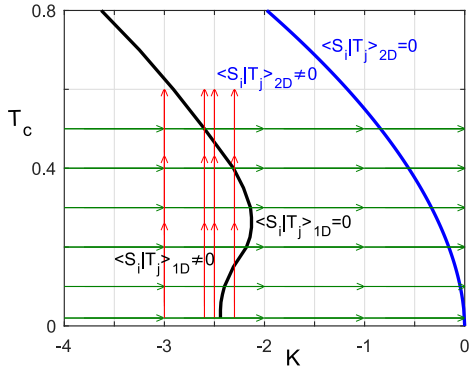}
\caption{\label{Fig3} One- and two-dimensional cases. The regions with zero and nonzero spin--pseudospin correlations. Black and blue solid lines denote the phase boundary in 1D and 2D cases, respectively. Straight lines with arrows are the paths, along which the results for the 1D case are represented (see Figs.~\ref{Fig4}--\ref{Fig6}).}
\end{center}
\end{figure}

Figure~\ref{Fig3} shows phase diagrams for both the 1D and 2D cases. We can see that the one-dimensional case is qualitatively different from the two-dimensional one. Indeed, in contrast to the 2D case, the line separating the regions with zero and nonzero spin--pseudospin correlations in the 1D case begins at a nonzero intersubsystem exchange coupling constant (at $K \approx -2.44$ and $T=0$). Moreover, the phase boundary is characterized by a nonmonotonic behavior suggesting the possibility of a reentrant transition to the region with entanglement under variation of the temperature.

\begin{figure}[tp]
\begin{center}
\includegraphics[width=0.8\columnwidth]{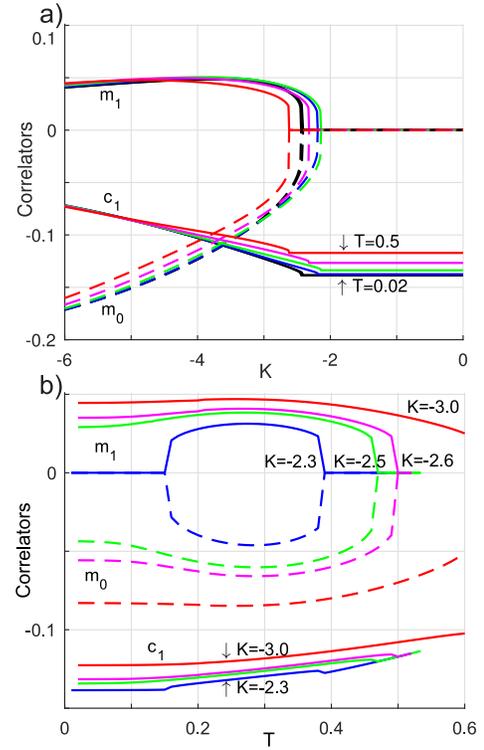}
\caption{\label{Fig4} One-dimensional lattice. (a) Correlation functions versus the intersubsystem exchange coupling constant $K$ at different temperatures $T$  given in different colors. (b) Temperature dependence of the correlation functions at indicated exchange coupling constants $K$ given in different colors. Dashed, upper solid, and lower solid lines denote the on-site spin--pseudospin correlation function $m_{0}$, the spin--pseudospin correlation function $m_{1}$  at the nearest-neighbor sites, and the spin--spin correlation function $c_{1}$ at the nearest-neighbor sites, respectively.}
\end{center}
\end{figure}

Of course, this nonmonotonicity manifests itself in other results. In Figs. \ref{Fig4}a and \ref{Fig4}b for the 1D chain (analogous to Figs.~\ref{FigS1} and \ref{FigS3} for the 2D case presented in Appendix), we illustrate the evolution of spin--spin and spin--pseudospin correlation functions changing along the horizontal and vertical paths shown in Fig.~\ref{Fig3}(as a function of $K$  at fixed $T$ and as a function of $T$ at fixed $K$).

The plots in Fig.~\ref{Fig4} are not as pictorial as similar ones for the 2D case shown in Fig.~\ref{FigS1} because the curves corresponding to the temperatures at both sides from the inflection point in the phase boundary overlap. An additional qualitative difference from the 2D case is a kink in the spin--spin correlation function $c_{1}$ arising at the $K_c$ point simultaneously with the beginning of the power-law growth of on-site ($m_{0}$) and intersite ($m_{1}$) spin--pseudospin correlation functions.

The most clearly pronounced feature in Fig.~\ref{Fig4}b in the form of a ``bubble" at $K = -2.3$  is related just to the reentrant transition (at entering the temperature range of entanglement and leaving it). In Fig.~\ref{Fig4}a, we can also see that the spin--spin correlation function $c_{1}$  exhibits a kink at the $K_c$ point and decreases steeply in absolute value with the increase in $|K|$  within the range of entanglement simultaneously with an increase in the spin--pseudospin correlation functions.

In the 2D case, only the intersubsystem correlations exhibit an anomalous behavior. Therefore, it is intuitively obvious that the simultaneous existence of anomalous features in both intra- and intersubsystem types of correlation functions in the 1D case can underlie substantial difference between these two cases, at least, in the behavior of the specific heat.

\begin{figure} [tp]
\begin{center}
\includegraphics[width=0.8\columnwidth]{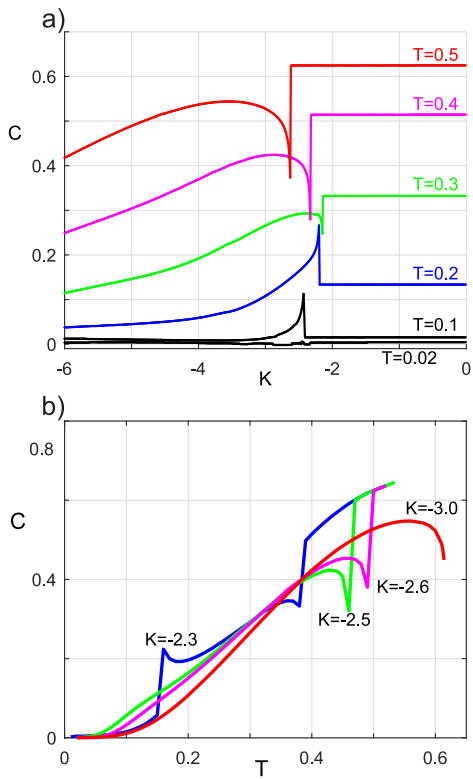}
\caption{\label{Fig5} One-dimensional lattice. (a) Specific heat versus the intersubsystem exchange coupling constant $K$ at indicated temperatures $T$  given in different colors. (b) Temperature dependence of the specific heat at indicated exchange coupling constants $K$  given in different colors. Compare this figure to Fig.~\ref{Fig1}.}
\end{center}
\end{figure}

Indeed, this is seen in Fig.~\ref{Fig5} (analogous to Fig~\ref{Fig1} for the 2D case). Here, as well as in the two-dimensional case, we see a jump in the specific heat at entering the range of entanglement. However, in contrast to the two-dimensional case, the sign of such a jump depends on the transition temperature. This jump is positive and negative in the lower and upper part of the 1D phase boundary (see Fig.~\ref{Fig3}). Such a behavior is inconsistent with the usual concepts of the specific heat jump accompanying transitions to a more ordered state and can be attributed to the interplay between different-sign changes in intra- and intersubsystem correlations. The fast increase in the spin--pseudospin correlation functions dominates over the decrease in spin--spin correlation functions at low temperatures and vice versa at temperatures above that corresponding to the inflection point at the phase boundary.

In the 1D case, the asymptotic behaviors of the specific heat plots in the limits of high and low temperatures are the qualitatively the same as in 2D at all values of $K$: all curves at high temperatures tend to a common asymptotic curve and $C \to 0$ for all plots at $T \to 0$.

\begin{figure}[t]
\begin{center}
\includegraphics[width=0.8\columnwidth]{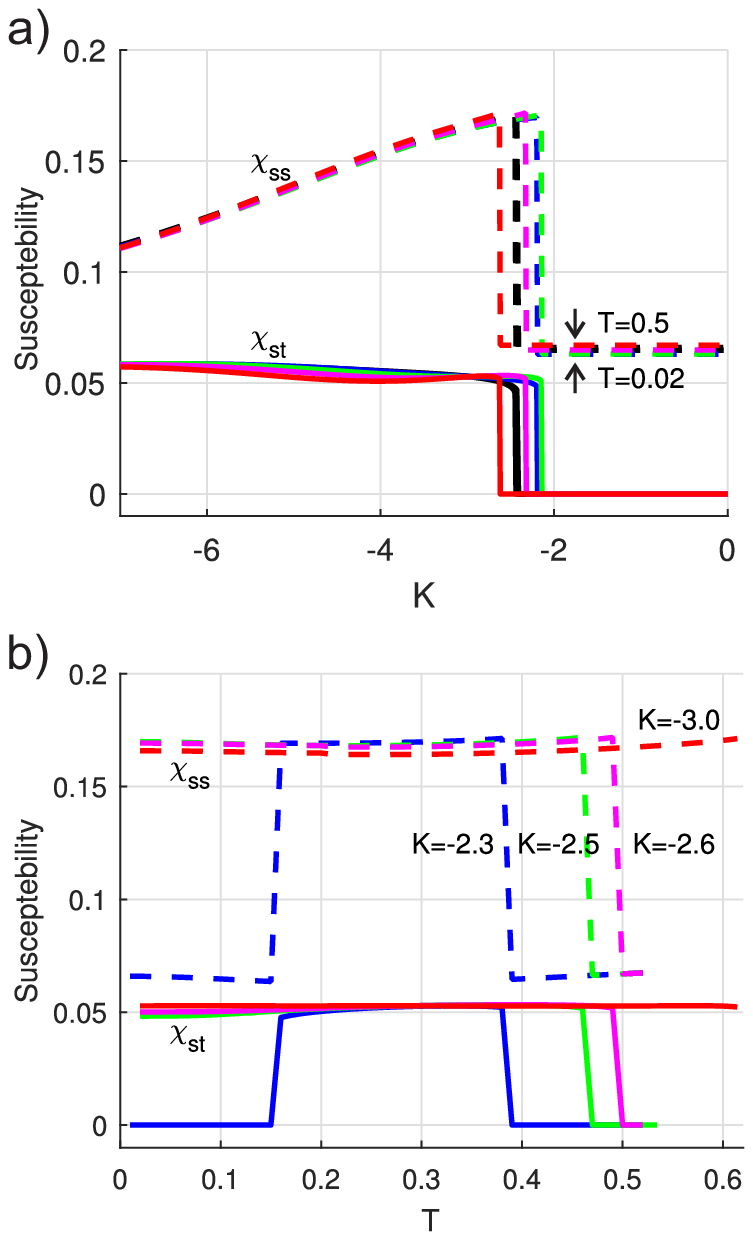}
\caption{\label{Fig6} One-dimensional lattice. (a) Susceptibility versus the intersubsystem exchange coupling constant $K$  at different temperatures $T$  given in different colors. (b) Temperature dependence of the susceptibility at indicated intersubsystem exchange coupling constants $K$ given in different colors. In both panels, dashed and solid lines denote the spin--spin susceptibility $\chi_{ss}$ and spin--pseudospin susceptibility $\chi_{ss}$, respectively. Compare this figure to Fig.~\ref{Fig2}.}
\end{center}
\end{figure}

Finally, in Fig.~\ref{Fig6} (analogous to Fig.~\ref{Fig2} for the 2D case), we illustrate the behavior of spin--spin $\chi_{ss}$ and spin--pseudospin $\chi_{st}$ susceptibilities along the horizontal and vertical paths shown in Fig.~\ref{Fig3}. The plots in Fig.~\ref{Fig6}a, as those in the similar figure for the correlation functions, are not pictorial enough because of the overlap of curves corresponding to the temperatures lying on both sides of the inflection point in the phase boundary. Nevertheless, we can see the fast decrease in the spin--spin correlation function with increasing $|K|$.

The two-side steps in Fig~\ref{Fig6}b, as well as the ``bubble" for the correlation function $m_1$ shown in Fig.~\ref{Fig4}b, are obviously due to the temperature-induced reentrant transition to the entanglement range and back.

In the rest aspects, the behavior of both susceptibilities is qualitatively the same as that in the two-dimensional case. Both the spin--spin $\chi_{ss}$  and spin--pseudospin  $\chi_{st}$ susceptibilities exhibit a jump at the critical points.

\section{Conclusions}
\label{concl}

The results obtained in this work clearly demonstrate the possibility of a transition to the state with quantum entanglement in spin--orbital systems even in the absence of long-range order. Such a transition characterized by the threshold onset of spin--orbital correlations is clearly similar to usual phase transitions (which manifests themselves in the features of the specific heat and susceptibilities). In fact, it is a phase transition in some quantum fluid. Therefore, such phenomena can be expected in other systems with the quantum entanglement. At the chosen signs of the model parameters ($I, J > 0$, $K < 0$), the entangled state occurs in wide ranges of the temperature and model parameters. The possibility of quantum entanglement in spin--orbital models is widely discussed in the literature (see, e.g., Refs.~\onlinecite{Oles12_JPCM,Lundgr12_PRB,Pati98_PRL,Itoi00_PRB,Chen07_PRB})) but usually at other relations between the parameters (in particular, when $K > 0$  at the same signs of $I$ and $J$). Such choice gives rise to entanglement only within a rather limited region of the phase diagram. We emphasize that the studied features of the thermodynamic characteristics seem to be of a purely quantum nature. Indeed, the analysis of the exact solution of the one-dimensional spin--orbital model for the case of classical spins~\cite{Kugel80_FNT} does not reveal the same features in the behavior of the spin--orbital correlation function (its sharp vanishing at certain values of the temperature and model parameters) and the corresponding peculiarities in the thermodynamic characteristics.

\section*{Acknowledgments}

This work was supported by the Russian Foundation for Basic Research, project nos. 17-52-53014 and 19-02-00509. K.I. Kugel also acknowledges the support of the Russian Foundation for Basic Research, project nos. 17-02-00323 and 17-02-00135.

\appendix
\setcounter{figure}{0}   
\renewcommand{\thefigure}{B\arabic{figure}} 
\section{Expressions for the Green's functions}

\subsection{General expressions}

Spin-spin and spin--pseudospin retarded Green's function are
\begin{equation}
G_{\mathbf{q}} = \left\langle S_{\mathbf{q}}^{z}\mid
S_{\mathbf{-q}}^{z}\right\rangle _{\omega } , \label{Gq0}
\end{equation}
\begin{equation}
R_{\mathbf{q}} = \left\langle T_{\mathbf{q}}^{z}\mid
S_{\mathbf{-q}}^{z}\right\rangle _{\omega } , \label{Rq0}
\end{equation}
Obviously, we have
$\left\langle T_{\mathbf{q}}^{z}\mid T_{\mathbf{-q}}^{z}\right\rangle _{\omega} = \left\langle S_{\mathbf{q}}^{z}\mid S_{\mathbf{-q}}^{z}\right\rangle_{\omega}$,
because we consider symmetric case $I=J$.

The self-consistent spherically symmetric approach leads to the following expressions for $G_{\mathbf{\mathbf{q}}}$ and $R_{\mathbf{q}}$:
\begin{equation}
G_{\mathbf{q}} = \frac{F_{ac}(\mathbf{q})}{\omega ^{2}-\omega
_{ac}^{2}(\mathbf{q})}+\frac{F_{opt}(\mathbf{q})}{\omega ^{2}-\omega
_{opt}^{2}(\mathbf{q})} , \label{Gq_A}
\end{equation}
\begin{equation}
R_{\mathbf{q}} = \frac{F_{ac}(\mathbf{q})}{\omega ^{2}-\omega
_{ac}^{2}(\mathbf{q})}-\frac{F_{opt}(\mathbf{q})}{\omega ^{2}-\omega
_{opt}^{2}(\mathbf{q})} , \label{Rq_A}
\end{equation}

\subsection{2D case}

The numerators for acoustic and optical branches are
\begin{equation}
F_{ac}=\frac{F_{1}+F_{2}}{2},\ F_{opt}=\frac{F_{1}-F_{2}}{2},
\end{equation} \begin{equation}
F_{1}=-8Jc_{g}(1-\gamma _{\mathbf{q}})-Mm_{0}, \ F_{2}=Mm_{0}, \label{F12}
\end{equation}
and the excitations spectra
\begin{equation}
\omega_{ac}^{2}(\mathbf{q})= W_1 + W_2,\ \omega _{opt}^{2}(\mathbf{q})= W_1 - W_2, \label{spectra}
\end{equation} \begin{eqnarray}
W_1&=&2J^{2}(1\!-\!\gamma _{\mathbf{q}})\Bigl\{1
+4\bigl[\widetilde{c}_{2g}\!+\!2\widetilde{c}_{d}\!-\!\widetilde{c}_{g}(1\!+\!4\gamma _{\mathbf{q}})\bigr]\Bigr\}\!+\notag \\
&&+4JM(2\widetilde{m}_{g}-\widetilde{m}_{g}\gamma_{\mathbf{q}}-\widetilde{m}_{0}\gamma _{\mathbf{q}})+\frac{1}{8}M^{2}, \label{W_1} \\
W_2&=&-4JM\left[ \widetilde{c}_{g}(1-\gamma _{\mathbf{q}})+\widetilde{m}_{g}-\widetilde{m}_{0} \gamma_{\mathbf{q}}\right]
 -\frac{1}{8}M^{2}, \label{W_2}
\end{eqnarray}
here $c_{r}=\langle \widehat{S}_{\mathbf{i}}^{z}\widehat{S}_{\mathbf{i+r}}^{z}\rangle,
\quad r = g, d, 2g$
 are spin-spin correlation functions, respectively for first (side of the square) $c_{g} \equiv c_{1}$, second (diagonal) $c_{d} \equiv c_{2}$ and third (doubled side) $c_{2g} \equiv c_{3}$ nearest neighbors, $\widetilde{c}_{r} = \alpha_{r} {c}_{r}$ are correlation functions with vertex corrections. The lattice sum for square case $\gamma_{\mathbf{q}}=\frac{1}{4}\sum_{\mathbf{g}}e^{i\mathbf{qg}}=
\frac{1}{2}(\cos (q_{x})+\cos (q_{y}))$.

Hereinbefore on-site $m_{0}$ and intersite $m_{g} \equiv m_{1}$ spin-pseudospin correlation functions are
\begin{equation}
m_{0}=\langle S_{\mathbf{i}}^{z}T_{\mathbf{i}}^{z}\rangle,\ \
m_{g}=\langle S_{\mathbf{i}}^{z}T_{\mathbf{i+g}}^{z}\rangle ,
\end{equation}
and for the intersubsystem vertex corrections
$\widetilde{m}_{0}=\alpha _{ST}^{0}m_{0}$,
$\widetilde{m}_{g}=\alpha _{ST}^{g}m_{g}$
we adopted the approximation
$\alpha _{ST}^{0}=\alpha _{ST}^{g}=1$.
For simplicity, we use the notation $M=8Km_{0}$.

Note, that the following relations for the symmetrical points $\mathbf{\Gamma}=(0,0)$ and $\mathbf{Q}=(\pi,\pi)$ in the Brillouin zone are always satisfied
\begin{equation}
\omega_{opt}(\mathbf{\Gamma})\geq\omega_{ac}(\mathbf{\Gamma})=0, \quad
\omega_{ac}(\mathbf{Q})\geq\omega_{opt}(\mathbf{Q})\geq 0. \label{omegas}
\end{equation}

\subsection{1D case.}
\begin{equation}
F_{ac}=\frac{F_{1}+F_{2}}{2},\ F_{opt}=\frac{F_{1}-F_{2}}{2},
\end{equation}
\begin{equation}F_{1}=-4Jc_{g}(1-\gamma _{\mathbf{q}})-Mm_{0}, \ F_{2}=Mm_{0} \label{F12_1D}
\end{equation}
\begin{eqnarray}
W_1&=&J^{2}(1-\gamma _{\mathbf{q}})\Bigl\{ 1+4\bigl[\widetilde{c}_{2g}-
\widetilde{c}_{g}(1+2\gamma _{\mathbf{q}})\bigr] \Bigr\} + \notag \\
&&2JM\left( 2\widetilde{m}_{g}-\widetilde{m}_{o}
\gamma_{\mathbf{q}}-\widetilde{m}_{g}\gamma _{\mathbf{q}}\right) +\frac{1}{8}M^{2}, \label{W_1} \\
W_2&=&-2JM\left[ \widetilde{c}_{g}(1-\gamma _{\mathbf{q}})+\widetilde{m}%
_{g}-\widetilde{m}_{0}\gamma _{\mathbf{q}}\right] -\frac{1}{8}M^{2}, \label{W_2}
\end{eqnarray}
now $M=4Km_{0}$, $\gamma_{\mathbf{q}}$ is one-dimensional, other notations are the same as for 2D case.

\newpage
\section{Three figures from Ref.~10}

\begin{figure}[H]
\begin{center}
\includegraphics[width=0.8\columnwidth]{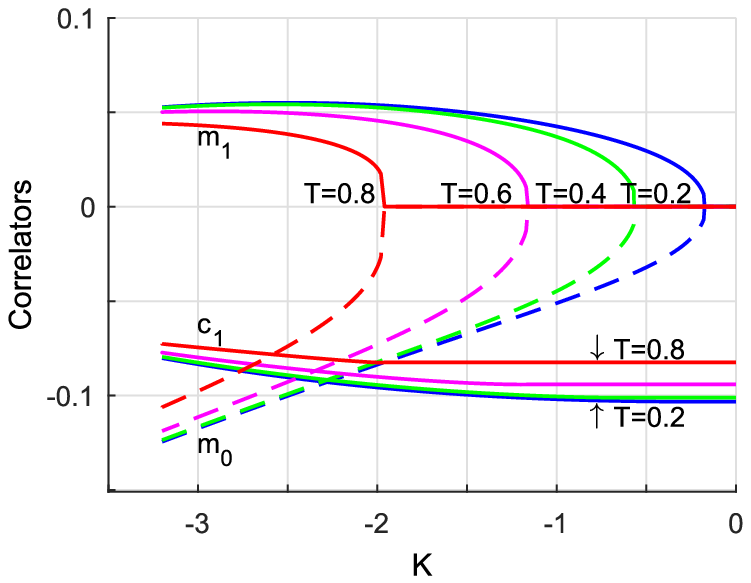}
\caption{\label{FigS1} 2D lattice.
Spin--spin and spin--pseudospin correlation functions versus the intersubsystem exchange parameter $K$ for different temperatures.
Spin--spin nearest neighbor correlation function $c_{1}$ --- lower solid lines, on-site spin--pseudospin correlation functions $m_{0}$ --- upper solid, and nearest neighbor spin--pseudospin correlation function $m_{1}$ --- dotted. Different colors correspond to different temperatures. For $m_{0}$ and $m_{1}$ curves, the temperatures are indicated at zero of the $y$ axis. For $c_{1}$ lines the boundary values of $T$ are indicated (from Ref.~\onlinecite{Kaga14_JL_R}).
}
\end{center}
\end{figure}
\begin{figure}[H]
\begin{center}
\includegraphics[width=0.8\columnwidth]{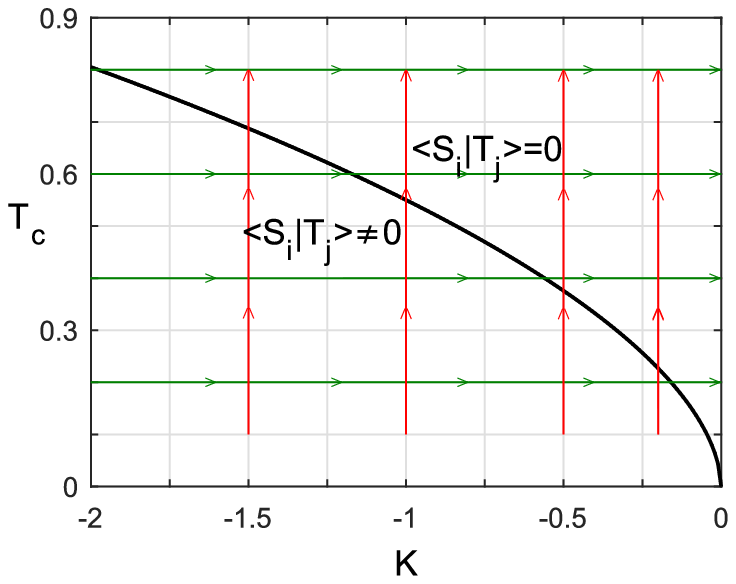}
\caption{\label{FigS2} 2D lattice.
Regions corresponding to zero and nonzero spin--pseudospin correlations. The phase boundary is well fitted by the $T_c=0.55|K|^{0.55}$ curve (from Ref.~\onlinecite{Kaga14_JL_R}). Lines with arrows show the paths, along with the results for 2D case are presented (see Figs.~\ref{FigS1} and \ref{FigS3}, as well as Figs.~\ref{Fig1} and \ref{Fig2} in the main text).
}
\end{center}
\end{figure}

\begin{figure}[H]
\begin{center}
\includegraphics[width=0.8\columnwidth]{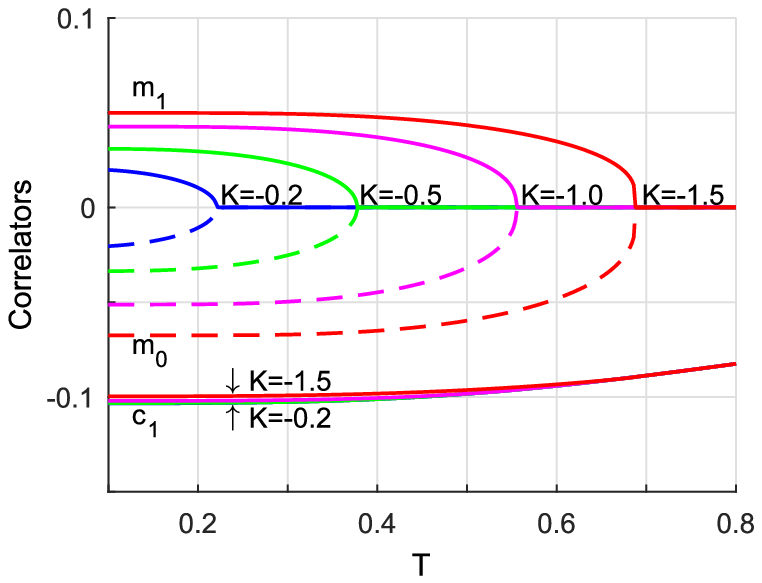}
\caption{\label{FigS3} 2D lattice.
Temperature dependence of the correlation functions at several fixed $K$ values. As in Fig.~\ref{FigS1}, spin--spin nearest neighbor correlation function $c_{1}$ --- lower solid lines, on-site spin--pseudospin correlation function $m_{0}$ --- upper solid, and nearest neighbor spin--pseudospin correlation functions $m_{1}$ --- dotted.
Different colors correspond to different $K$ values. For $m_{0}$ and $m_{1}$ curves, $K$ values are indicated at zero of the $y$ axis. For $c_{1}$ lines, the boundary $K$ values are indicated.
The curves $|m_{0}|(T)$ and $m_{g}(T)$ are well fitted by power law
$m\sim (T_c-T)^\alpha$ with the exponent $\alpha \sim 0.3--0.5$ nearly independent of $K$ (from Ref.~\onlinecite{Kaga14_JL_R}).
}
\end{center}
\end{figure}

In Figs.~\ref{FigS1}--\ref{FigS3} almost invisible nonzero values of $m_{0}$ and $m_{1}$ just before the transition ($T\gtrsim T_c$ for
Fig.~\ref{FigS1} and $T\gtrsim T_c$ for Fig.~\ref{FigS2}) show the accuracy of the self-consistent calculations.

\end{document}